\magnification 1200
\font\biga=cmbx10 scaled\magstep3
\font\sfa=cmr9
\hsize 16.5 true cm
\def\d${ $\displaystyle } 
\rightline{\bf DFUB 96-07}
\rightline{\bf TP-USL/96/01}
\rightline{ February 1996/Revised April 1996}
\vskip 1.0 cm

\centerline{\biga Simulation of the process $e^+ e^- \mapsto e^+ e^- \gamma$ }
\centerline{\biga within electroweak theory }
\centerline{\biga with longitudinally polarized initial electrons.}
\vskip 1.0 cm

\centerline{\bf 
M.~Caffo\ $^{ab}$, 
H.~Czy{\.z}\ $^c${\footnote 
{ $^{\star}$}{\sfa 
     Partly supported by 
     the Polish Committee for Scientific Research
     under grant no 2P03B17708 
     and
     USA-Poland Maria Sk{\l}odowska-Curie
     Joint Fund II under grant MEN/NSF-93-145
     . } },
E.~Remiddi\ $^{ba}$ }

\vskip 1.0 cm
\item{$^a$ \ }{\it INFN, Sezione di Bologna, I-40126 Bologna, Italy }
\item{$^b$ \ }{\it Dipartimento di Fisica, Universit\`a di Bologna, 
             I-40126 Bologna, Italy }
\item{$^c$ \ }{\it Institute of Physics, University of Silesia, 
             PL-40007 Katowice, Poland }

\par
\bigskip

\noindent
{\tt E-mail: \par
caffo@bologna.infn.it \par
czyz@usctoux1.cto.us.edu.pl \par
remiddi@bologna.infn.it \par} 
\bigskip

\vskip 1cm 
\centerline{\bf Abstract }
\par
We present simple analytic expressions for the distributions of the Bhabha 
scattering process with emission of one hard photon, including weak boson 
exchanges, and with longitudinal polarization of the initial electron. 
The results from the Monte Carlo generator BHAGEN-1PH, based on these 
expressions, are presented and compared, for the unpolarized case, with those 
existing in literature.

\vfill
\eject

\noindent
1. {\bf Introduction.}
\par
The cross section for $e^+ e^- \mapsto e^+ e^- \gamma$ in QED has been 
calculated for different purposes and various distributions have been produced 
[1-16]. 
The availability of high energies (TRISTAN, LEP, SLC) and longitudinally 
polarized electron beam (SLC) makes it necessary to include weak-boson and 
polarization effects.
For the unpolarized case a relatively simple expression for the square of the 
matrix element was obtained [2,3], for the polarized case it was advocated 
[4] that it is easier to use just a matrix element and not its square. 
This is true for the general case with multiphoton emission, however for the 
experimentally interesting case of longitudinally polarized electron beam, 
by careful choice of the variables, in [5] was obtained a reasonably compact
expression for the square of the matrix element. 
In this work that expression is improved with some other terms relevant on the 
$Z$ boson peak and with all the relevant mass-corrections for the 
configurations in which the final fermions have angles with the initial 
direction larger than, say, 1 mrad.
For extremely forward final fermions the mass-corrections reported here are 
not sufficient as discussed in [1,5,6,7,8].
We present the complete analytical form of the differential cross-section, 
omitting the details of the calculation; the expression is simple enough and 
can be easily cross-checked. 
Because of the experimental interest we also implemented it within a 
Monte Carlo program, allowing for a fast event generation of the process. 
Furthermore the procedure do not require the introduction of new peaks as 
frequently occurring in spinor techniques.
Some results and the numerical tests are given in Section 3, while 
the details of the structure of the program are presented elsewhere [9]. 
\bigskip

\noindent
2. {\bf The distributions.}
\par
We consider the process
 $$ e^+ (p_+ ) + e^- (p_- ) \rightarrow e^+ (q_+ ) + e^- (q_- ) 
    +  \gamma (k) . \leqno (2.1)$$
We define
$$ \eqalign { 
     s  &= (p_+ + p_-)^2  \ , \cr
     t_1 &= (p_- - q_-)^2  \ , \cr }
\eqalign { 
     t  &= (p_+ - q_+)^2  \ , \cr
     u_1 &= (p_- - q_+)^2  \ , \cr }
\eqalign { 
     u  &= (p_+ - q_-)^2  \ , \cr
     k_{\pm} &= p_{\pm}.k  \ , \cr }
\eqalign { 
     s_1 &= (q_+ + q_-)^2  \ , \cr
     h_{\pm} &= q_{\pm}.k  \ , \cr }
   \leqno(2.2) $$
and
$$ {\cal P}   = \epsilon_{\mu \nu \rho \sigma} 
  p_+^{\mu} p_-^{\nu} q_+^{\rho} q_-^{\sigma} \ ,
\qquad \qquad   \epsilon_{0123} = 1 \ .
\leqno(2.3) $$
In the following $\alpha$ is the fine structure constant, $P_L$
is the degree of longitudinal polarization of the initial electron
($P_L  = 1$ for pure right-handed states) and $C_V, C_A$ the $Z$ couplings
to the electron. 
In the standard electroweak theory one has in particular
$$ C_V =-{{1-4 \sin^2 \theta_W }\over{4 \sin \theta_W  \cos \theta_W }} ,
\qquad   C_A =-{{1}\over{4 \sin \theta_W  \cos \theta_W }}, 
\leqno(2.4)$$
where $\theta_W$ is the weak mixing angle.
We define, for later convenience, also the combinations 
(the factor 2 in $C_{ZZ}^0$ was absent in [5], due to a misprint)
$$ \eqalign {
              C_{\gamma Z}^{\pm}   &= C_V^2  \pm  C_A^2 \ , \qquad \cr   
              C_{\gamma Z}^{0}     &= C_V  C_A \  , \cr }
   \eqalign {
      C_{ZZ}^{\pm}   &=(C_V^2  + C_A^2  )^2 \pm 4 C_V^2   C_A^2 \ , \cr
      C_{ZZ}^0       &=2 C_V C_A (C_V^2  + C_A^2  ) \ . \cr}
   \leqno(2.5) $$
With $M_Z$ and $\Gamma_Z$ the mass and width of the $Z$, $\Theta$ the usual
step function  and
$$ D(s)= (s-M_Z )^2  + M_Z^2 \Gamma_Z^2 \Theta(s) \ , \leqno(2.6)$$
we define the following functions
(slightly different from [5], to keep the expressions as compact as possible)
 $$ \eqalign {
    F^+ (s,t,a)&= 1 + \left[ {{s(s-M_Z^2)}\over{D(s)}}
   + {{t(t-M_Z^2)}\over{D(t)}}\right]
   \left( C_{\gamma Z}^+ - 2 P_L C_{\gamma Z}^0(1+a) \right) \cr
   &+ {{st\left[(s-M_Z^2)(t-M_Z^2) 
      + M_Z^2 \Gamma_Z^2 \Theta(s) \Theta(t)\right]}\over
     {D(s) D(t)} } 
   \left(C_{ZZ}^+ - 2 P_L C_{ZZ}^0(1+a) \right) \ , \cr }  \leqno(2.7) $$
 $$ \eqalign { 
   F^- (s,t)&= 1 + \left[ {{s(s-M_Z^2)}\over{D(s)}}
   + {{t(t-M_Z^2)}\over{D(t)}}\right]
    C_{\gamma Z}^- \cr 
   &+ {{st\left[(s-M_Z^2)(t-M_Z^2) 
      + M_Z^2 \Gamma_Z^2 \Theta(s) \Theta(t)\right]}\over
     {D(s) D(t)} } 
   C_{ZZ}^- \ , \cr }    \leqno(2.8) $$
 $$ \eqalign{
   E^+(s,t) &= M_Z \Gamma_Z \biggl[
   \left( {{s\Theta(s)}\over{D(s)}} - {{t\Theta(t)}\over{D(t)}}\right)
   \left(C_{\gamma Z}^0 - {1\over 2} P_L C_{\gamma Z}^+ \right) \cr
  &+ {{st\left( \Theta(s)(t-M_Z^2) - \Theta(t)(s-M_Z^2) \right)}
     \over {D(s) D(t)}}
   \left( C_{ZZ}^0 - {1\over 2} P_L C_{ZZ}^+ \right)
    \biggr] \ , \cr } \leqno(2.9) $$
 $$ \eqalign { 
   E^-(s,t) &= - {{P_L}\over 2}  M_Z \Gamma_Z \biggl[
   \left( {{s\Theta(s)}\over{D(s)}} - {{t\Theta(t)}\over{D(t)}}\right)
    C_{\gamma Z}^- \cr
  &+ {{st\left( \Theta(s)(t-M_Z^2) - \Theta(t)(s-M_Z^2) \right)}
     \over {D(s) D(t)}}
    C_{ZZ}^- 
    \biggr] \ . \cr } \leqno(2.10) $$
The differential cross section for the process (2.1) can be written as
 $$ d\sigma  = {{\alpha^3}\over{2\pi^2 s}} (X+Y+Z) {{d^3q_+}\over{E_+}}
      {{d^3q_-}\over{E_-}} {{d^3k}\over{E_{\gamma}}}
       \delta^4(p_+ + p_- - q_+ - q_- - k) \ , \leqno(2.11) $$
where $E_+, E_-, E_{\gamma}$ are the energies of the final positron, 
electron and photon respectively. 
The quantities $X, Y, Z$  refer to the annihilation, the coulomb and the
interference part of the square amplitude respectively and, omitting the 
terms which are numerically irrelevant at the interested energies and for 
the kinematical configurations considered (final fermions not too forward), 
can be expressed in the form
$$ \eqalign { 
 X =
   &\left[ F^- (s,s_1)  (t^2 +t_1^2 ) + F^+ (s,s_1,0)  (u^2 +u_1^2 )\right] 
    {1\over{4 s s_1}}\left[{u\over{k_+h_-}} +{u_1\over{k_-h_+}} 
                         -{t\over{k_+h_+}} -{t_1\over{k_-h_-}} \right]  \cr
  +&\left[ F^- (s_1,s_1) (t^2 +t_1^2 ) + F^+ (s_1,s_1,0) (u^2 +u_1^2 )\right]
    {1\over{4 s_1 k_+  k_-}}   \cr
  +&\left[ F^- (s,s)   (t^2 +t_1^2 ) + F^+ (s,s,0)   (u^2 +u_1^2 )\right]
    {1\over{4 s  h_+ h_-}}  \cr
  +& {\cal P} \left[ E^-(s,s_1) (t^2 -t_1^2) + E^+(s,s_1) (u^2 -u_1^2)  \right]
       {{s-s_1}\over{2 s s_1 k_+ k_- h_+ h_-}} \cr
  -&{{m_e^2}\over{2 s^2}} \left[ 
     F^-(s,s) \left({{t_1^2}\over{(h_+)^2}} + {{t^2}\over{(h_-)^2}}\right) 
   + F^+(s,s,0) \left({{u^2}\over{(h_+)^2}} + {{u_1^2}\over{(h_-)^2}} \right) 
   \right] \cr
  -&{{m_e^2}\over{2 s_1^2}} \left[ 
     F^-(s_1,s_1) \left({{t_1^2}\over{(k_+)^2}} + {{t^2}\over{(k_-)^2}}\right) 
   + F^+(s_1,s_1,0) {{u_1^2}\over{(k_+)^2}} + F^+(s_1,s_1,a) {{u^2}\over{(k_-)^2}} 
   \right] \ , \cr
   } \leqno(2.12) $$

$$ \eqalign { 
 Y =
    &\left[ F^- (t,t_1) (s^2 +s_1^2 ) + F^+ (t,t_1,0) (u^2 +u_1^2 )\right]
      {1\over{4 t t_1}} 
     \left[{u\over{k_+h_-}} + {u_1\over{k_-h_+}} + {s\over{k_+k_-}}
        + {s_1\over{h_+h_-}} \right]  \cr
   -&\left[ F^- (t,t) (s^2 +s_1^2 ) + F^+ (t,t,0) (u^2 +u_1^2 )\right]
    {1\over{4 t k_- h_-}} \cr
   -&\left[ F^- (t_1,t_1) (s^2 +s_1^2 ) + F^+ (t_1,t_1,0) (u^2 +u_1^2 )\right]
    {1\over{4 t_1 k_+ h_+}} \cr
   -&{{m_e^2}\over{2 t_1^2}} \left[ 
     F^-(t_1,t_1) \left({{s^2}\over{(h_+)^2}} + {{s_1^2}\over{(k_+)^2}}\right) 
   + F^+(t_1,t_1,0) \left({{u^2}\over{(h_+)^2}} + {{u_1^2}\over{(k_+)^2}} \right) 
     \right] \cr
   -&{{m_e^2}\over{2 t^2}} \left[ 
     F^-(t,t) \left({{s^2}\over{(h_-)^2}} + {{s_1^2}\over{(k_-)^2}}\right) 
   + F^+(t,t,0) {{u_1^2}\over{(h_-)^2}} 
   + F^+(t,t,a) {{u^2}\over{(k_-)^2}} 
     \right] \ , \cr
   } \leqno(2.13) $$

$$ \eqalign { 
 Z = &{{u^2 + u_1^2}\over 4} \cr
    \biggl[ &{{F^+(s,t,0)}\over{st}} 
     \left( {{u}\over {k_- h_+}} + {{s}\over{h_+h_-}} 
            - {{t}\over{k_-h_-}} \right) +
   {{F^+(s,t_1,0)}\over{st_1}} 
     \left( {{u_1}\over {k_+ h_-}} + {{s}\over{h_+h_-}} 
            - {{t_1}\over{k_+h_+}} \right) \cr
           +&{{F^+(s_1,t,0)}\over{s_1t}} 
     \left( {{u_1}\over {k_+ h_-}} + {{s_1}\over{k_+k_-}} 
            - {{t}\over{k_-h_-}} \right) +
   {{F^+(s_1,t_1,0)}\over{s_1t_1}} 
     \left( {{u}\over {k_- h_+}} + {{s_1}\over{k_+k_-}} 
            - {{t_1}\over{k_+h_+}} \right) \biggr] \cr
+& {\cal P} \left[ {{E^+(s,t)k_+}\over{st}} + {{E^+(s,t_1)k_-}\over{st_1}}
            - {{E^+(s_1,t)h_+}\over{s_1t}} - {{E^+(s_1,t_1)h_-}\over{s_1t_1}}
         \right] {{u^2-u_1^2}\over{k_+ k_- h_+ h_-}} \cr
   -&{{m_e^2}\over{ s t_1}}  F^+(s,t_1,0) {{u^2}\over{(h_+)^2}} 
    -{{m_e^2}\over{ s_1 t_1}} F^+(s_1,t_1,0) {{u_1^2}\over{(k_+)^2}}  \cr
   -&{{m_e^2}\over{ s t}}  F^+(s,t,0) {{u_1^2}\over{(h_-)^2}} 
    -{{m_e^2}\over{ s_1 t}} F^+(s_1,t,a) {{u^2}\over{(k_-)^2}} \ . \cr 
 } \leqno(2.14) $$  
where $m_e$ is the electron mass and $a = {{(s-s_1)^2}\over{ss_1}}$.
The mass-terms (i.e. the terms proportional to $m_e^2$) included here 
are the relevant ones when the final fermions have angles with the initial 
direction larger than, say, 1 mrad. 
For even smaller angles a few more terms have to be included as discussed 
in [1,5,6,7,8].
The contributions to Eq.s (2.12-14) proportional to the pseudoscalar
of Eq. (2.3), not reported in [5], are relevant at the $Z$ boson peak.
The unpolarized cross section can be easily recovered by putting $P_L=0$.
\par
We have verified that the unpolarized annihilation part agrees with the result
of Ref.[17] for $e^+ e^- \mapsto \mu^+ \mu^- \gamma$, in the limit of 
massless fermions. 
The  coulomb  part can  be  verified  by  noting  that  it  must be 
reproduced by crossing from the annihilation, i.e. by the exchange 
of  $p_+$ with $-q_-$. Finally, after some algebra, one can recover 
from our expressions the QED result of Ref.[2] in the appropriate limit; 
as a consequence of the presence of the $Z$-exchange terms, our formulae 
do not  posses the factorization observed there. 
Such a factorization is recovered only in the
unrealistic $M_Z=0$ limit, suggesting that the connection between hard and soft
photon emission observed there is related to the absence of a mass scale.
Our results reproduce also the results of Ref. [3] for the unpolarized case 
(in the t-channel $Z$ boson contribution we put to zero the absorptive part, 
i.e. the part proportional to the $Z$ width).
\bigskip
\noindent
3. {\bf Numerical results and comparisons.}
\par
On the basis of the above formulae we have constructed a Monte Carlo
event generator, BHAGEN-1PH. The algorithms used in the generation and
a long write-up can be found in Ref. [9].
However an indispensable part of the construction 
of the Monte Carlo event generator is its test, which assure that the 
algorithms used for the generation are well constructed and applied. 
Numerical tests check also a number of bugs in the program. 
The tests through comparisons with existing independent programs where 
possible only for the situation where the beams are unpolarized
as no program which can provide cross section for polarized beam(s) 
is available to our knowledge. 
However as the algorithm used in BHAGEN-1PH for generation in the situation
with the polarized electron beam is identical to the one used for unpolarized
beams, no further tests of the algorithm are necessary.
On the occasion of the estimation of the theoretical error of the luminosity 
measurement at LEP [10,11], extensive tests of the QED only 
$t$-channel part of the program were done, as BHAGEN-1PH was included as 
a part of BHAGEN95, a Monte Carlo program 
to calculate the cross-section for Bhabha scattering, completed with 
radiative corrections.
There the accuracy of $3 \times 10^{-4}$ was reached in the $\cal O(\alpha)$ 
comparisons; no deviation up to this accuracy was observed [10,11] from 
OLDBIS [12], SABSPV [13], NNLBHA [14], BHLUMI [15].
To complete this tests we present here a comparison with OLDBIS, where
only contributions to the cross section of the hard photon spectrum 
are separately compared in Table 1. 
This specific tests were not done in [10,11] as only the total error 
on the complete Bhabha scattering was relevant there.
The event selection used here is the BARE1 type [11] and the other cuts 
and parameters used are described in the caption of Table 1.
The results of the various programs agree again within their statistical 
errors, although the accuracy reached in this comparisons is much lower 
especially at the hard end of the spectrum, due to the long time necessary 
to run OLDBIS to obtain the required accuracy. 
\par
The program BHAGEN-1PH can be used of course also at lower energies, 
for example in the DA$\Phi$NE energy range, for which one  can
simulate the so called large-large angle electron-positron double tagging [6].
For the other extremely forward configurations described in [6] the formulae 
used in BHAGEN-1PH and presented in this paper are not adequate. 
In the Table 2 we show a comparisons of the results of BHAGEN-1PH with the 
results of PHIPHI, which were presented in [6] and [7]. 
The cuts used are described in the caption of Table 2, where the results 
are seen to agree well, although the large numerical errors.
Again the accuracy of the comparisons is limited by the accuracy
of the PHIPHI program results. 
As the energy is low these comparisons do not test the $Z$ contributions 
to the cross-section and again $t$-channel contributions are mostly concerned, 
due to the angular cuts used.
The tests are completed by the comparison with the BABAMC program [16]
for large angles around the $Z$ boson resonance, which allows for the
tests of $s$-channel $Z$ boson exchange contributions. 
As one can see from the Table 3 the agreement with BABAMC is good except 
for the hard part ($ 0.5 \ E_b < E_{\gamma} < 0.9 \ E_b $) of the photon 
spectrum at beam energy $E_b = 47.5$ GeV, where we have found the 
difference of 0.60(17)\% . 
We guess that this difference is due to the approximation
used in BABAMC to simulate this part of the spectrum, although
the accuracy seems to be sufficient for every expected experimental 
precision in this part of the allowed phase space.
Finally in Table 4 we present some cross-section results obtained with 
the longitudinally polarized initial electron, with beam degree of 
polarization $P_L =\pm 1$, in the experimentally 
accessible region $0.1 < E_{\gamma}/E_b<0.5$, cuts and 
parameters are specified in the table caption. 
\bigskip
\noindent
4.  {\bf Conclusions.}
\par
We have obtained simple analytic formulae for the cross-section of the 
$e^+ e^- \mapsto e^+ e^- \gamma$ process, with $Z$ exchange included and 
longitudinally polarized initial electron.
A fast Monte Carlo event generator simulating the process, BHAGEN-1PH,  
has been implemented to cover the experimental event selection, although 
in its present version the program is not suitable for configurations 
where final fermions are extremely forward ($\theta_{\pm} < 1 $mrad).
\par
Apart from the intrinsic interest for a direct independent test of the 
electroweak couplings, the bremsstrahlung process constitutes an important 
background to other processes.
\par
Together with the existing soft and collinear corrections the program is 
used [10,11] to calculate radiative corrections to the Bhabha scattering, 
including $Z$ exchange, for different detector arrangements.
\bigskip
\noindent
{\bf Acknowledgments.} 
\par
H.C. is grateful to the Bologna Section of INFN and to the Department 
of Physics of Bologna University for support and kind hospitality.
\vfill
\eject

\noindent 
{\bf References.} 
\par
\item{[1]\ } G. Altarelli and B. Stella, Lett. Nuovo Cim. 9 (1974) 416;  \par
             V.N. Baier, V.S. Fadin, V.A. Khoze and E.A. Kuraev, 
             Phys. Rep. 78 (1981) 293;                                   \par
             K.Tobimatsu and Y.Shimizu, Progr.Theor.Phys. 75 (1986) 905; \par
             M. Caffo, R. Gatto, E. Remiddi and F. Semeria, Nucl. Phys. 
             B327 (1989) 93.

\item{[2]\ } F.A. Berends et al., Phys. Lett. 103 B (1981) 124.

\item{[3]\ } F.A. Berends et al., Nucl. Phys. B206 (1982) 61.

\item{[4]\ } F.A. Berends et al., Nucl. Phys. B239 (1984)382; \par
             R. Kleiss, Zeit. f. Phys. C 33 (1987) 433.

\item{[5]\ } M. Caffo, R. Gatto and E. Remiddi, Nucl. Phys. B286 (1987) 293. 

\item{[6]\ } M. Greco et al., Phys. Lett B318 (1993) 635.

\item{[7]\ } G. Montagna, F. Piccinini and O. Nicrosini, Comp. Phys. Comm.
             78 (1993) 155; ibidem 79 (1994) 351.

\item{[8]\ } R. Kleiss, Phys. Lett. B318 (1993) 217;         \par
             R. Kleiss and H. Burkhardt, Com. Phys. Comm. 81 (1994) 372.
 
\item{[9]\ } M. Caffo and H. Czy{\.z}, in preparation.

\item{[10]\ } M. Caffo, H. Czy{\.z}, E. Remiddi, in CERN Yellow Report 
             CERN 95-03, D. Bardin, W. Hollik, G. Passarino Ed.s, p.361 (1995).

\item{[11]\ } H. Anlauf et al., in  CERN Yellow Report CERN 96-01, 
             G. Altarelli, T. Sj{\"o}strand and F. Zwirner Ed.s, (1996).

\item{[12]\ } F.A. Berends, R. Kleiss, Nucl. Phys. B228 (1983) 537.   \par
             S. Jadach et al., Phys. Lett. B253 (1991) 469.

\item{[13]\ } M. Cacciari et al., Comp. Phys. Commun. 90 (1995) 301.

\item{[14]\ } A. Arbuzov et al., in CERN Yellow Report CERN 95-03,
              D. Bardin, W. Hollik, G. Passarino Ed.s , p.369 (1995); 
              JETPh 108 (1995) 1.

\item{[15]\ } S. Jadach et al., Phys. Lett. B353 (1995) 362.

\item{[16]\ } F.A. Berends, R. Kleiss and W. Hollik, Nucl. Phys. B304 (1988)712.

\item{[17]\ } F.A. Berends, R. Kleiss and S. Jadach, Nucl. Phys. B202 (1982)63.
\par
\vfill\eject 
 
\def\crv{\cr \phantom{=} & & & & \cr \noalign{\hrule} }
\def\cru{\cr \phantom{=} & & & & \cr \noalign{\hrule}
         \cr \phantom{=} & & & & \cr }
 
\vbox{ \offinterlineskip \halign{
 \vrule\ # &\vrule\ # &\vrule\ # &\vrule\ # &             # \vrule \cr
\noalign{\hrule} \phantom{=} & & & & \cr 
  \d$ \Delta $
& OLDBIS \ \  
& BHAGEN-1PH
& BG/OB
& \cru
  0.01 - 0.1
& 12.235(4)
& 12.2374(5)
& 1.0002(4)
& \cru
  0.1 - 0.3
& 4.861(4)
& 4.8557(3)
& 0.9989(9)
& \cru
  0.3 - 0.5
& 1.631(4)
& 1.63236(15)
& 1.001(3)
& \cru
  0.5 - 0.7
& 0.630(4)
& 0.63183(9)
& 1.003(7)
& \cru
  0.7 - 0.9
& 0.303(4) 
& 0.31061(5)
& 1.025(14)
& \crv
  }}

\bigskip
{\bf Table 1.} \ 
Total cross-section (in nanobarns) are compared between 
OLDBIS [12] and BHAGEN-1PH 
for $t$-channel contribution to the $e^+ e^- \mapsto e^+ e^- \gamma$ process, 
for different ranges of the photon emitted energy 
($\Delta = {{E_{\gamma}}\over{E_b}}$), for the electron and positron 
angular range $3^\circ \leq \theta_\pm \leq 8^\circ $ and for beam energy
$E_b = $ 47.585 GeV. No cuts are applied on photon angles. 
The statistical variance (one standard deviation) is given in brackets 
In the third column is the ratio of BHAGEN-1PH cross-section to OLDBIS
cross-section. 

\vskip 2.0 cm
 
\def\crv{\cr \phantom{=} & & & & \cr \noalign{\hrule} }
\def\cru{\cr \phantom{=} & & & & \cr \noalign{\hrule}
         \cr \phantom{=} & & & & \cr }
 
\vbox{ \offinterlineskip \halign{
 \vrule\ # &\vrule\ # &\vrule\ # &\vrule\ # &             # \vrule \cr
\noalign{\hrule} \phantom{=} & & & & \cr 
  \d$ E_{\gamma}^{min} $ (MeV)
& PHIPHI \ \ 
& BHAGEN-1PH
& BG/PH
& \cru
  2
& 15.9(6)
& 16.18(7)
& 1.02(4)
& \cru
  5
& 13.0(5)
& 13.09(6)
& 1.01(5)
& \cru
  10
& 10.7(4)
& 10.78(5)
& 1.01(4)
& \cru
  50
& 5.66(21)
& 5.68(2)
& 1.00(4)
& \cru
  100
& 3.6(1) 
& 3.689(14)
& 1.02(3)
& \crv
  }}

\bigskip
{\bf Table 2.} \ 
Total cross-section (in microbarns) are compared between 
PHIPHI (values in [6] and correction factor in [7]) and BHAGEN-1PH 
for the $e^+ e^- \mapsto e^+ e^- \gamma$ process, for different
ranges of the photon emitted energy 
($ E_{\gamma}^{min} < E_{\gamma} < E_b-m_e^2/E_b$), 
for the electron and positron angular range 
$8.5^\circ \leq \theta_\pm \leq 171.5^\circ $ and for beam energy
$E_b = $ 0.51 GeV. The minimal allowed angle between photon and
final lepton is $0.5^\circ$.
The statistical variance (one standard deviation) is given in brackets. 
In the third column is the ratio of BHAGEN-1PH cross-section to PHIPHI
cross-section. 

\vfill
\eject 
 
\def\crv{\cr \phantom{=} & & & & & \cr \noalign{\hrule} }
\def\cru{\cr \phantom{=} & & & & & \cr \noalign{\hrule}
         \cr \phantom{=} & & & & & \cr }
 
\vbox{ \offinterlineskip \halign{
 \vrule\ # &\vrule\ # &\vrule\ # &\vrule\ # &\vrule\ # &     # \vrule \cr
\noalign{\hrule} \phantom{=} & & & & & \cr 
   \d$ E_b$ (GeV)
&  \d$ \Delta $
& BABAMC
& BHAGEN-1PH
& BG/BC
& \cru
  43.5
& 0.1 - 0.5
& 74.143(15)
& 74.150(5)
& 1.0001(3)
& \cru
  43.5
& 0.5 - 0.9
& 15.970(13)
& 15.9824(14)
& 1.0008(9)
& \cru
  45.5
& 0.1 - 0.5
& 241.24(8)
& 241.29(3)
& 1.0002(5)
& \cru
  45.5
& 0.5 - 0.9
& 55.34(7)
& 55.43(1)
& 1.0016(14)
& \cru
  47.5
& 0.1 - 0.5 
& 63.780(11)
& 63.758(5)
& 0.99966(25)
& \cru
  47.5
& 0.5 - 0.9 
& 8.325(13)
& 8.3750(8)
& 1.0060(17)
& \crv
  }}

\bigskip
{\bf Table 3.} \ 
Total cross-section (in picobarns) are compared between 
BABAMC [16] and BHAGEN-1PH 
for the $e^+ e^- \mapsto e^+ e^- \gamma$ process, for different
ranges of the photon emitted energy ($\Delta = {{E_{\gamma}}\over{E_b}}$)
and beam energy $E_b $, for the electron and positron angular range 
$40^\circ \leq \theta_\pm \leq 140^\circ $. 
No cuts are applied on photon angles. 
The statistical variance (one standard deviation) is given in brackets. 
In the third column is the ratio of BHAGEN-1PH cross-section to BABAMC
cross-section. The following parameters are used: $Z$ boson mass
 $M_Z = 91.175$ GeV, $Z$ boson width $\Gamma_Z = 2.3355$ GeV and 
 weak mixing angle $\sin\theta_W = 0.2247$.

\vskip 2.0 cm

 
\def\crv{\cr \phantom{=} & & & & & \cr \noalign{\hrule} }
\def\cru{\cr \phantom{=} & & & & & \cr \noalign{\hrule}
         \cr \phantom{=} & & & & & \cr }
 
\vbox{ \offinterlineskip \halign{
 \vrule\ # &\vrule\ # &\vrule\ # &\vrule\ # &\vrule\ # &     # \vrule \cr
\noalign{\hrule} \phantom{=} & & & & & \cr 
   \d$ E_b$ (GeV)
&  \d$ \Delta $
& $P_L =+1$
& $P_L = 0$
& $P_L =-1$
& \cru
  43.5
& 0.1 - 0.5
& 68.095(3)
& 74.150(4)
& 80.198(5)
& \cru
  45.5
& 0.1 - 0.5
& 200.90(1)
& 241.29(2)
& 281.69(2)
& \cru
  47.5
& 0.1 - 0.5 
& 57.469(3)
& 63.760(4)
& 70.062(5)
& \crv
  }}

\bigskip
{\bf Table 4.} \ 
Total cross-section (in picobarns) from BHAGEN-1PH for the process 
$e^+ e^- \mapsto e^+ e^- \gamma$, for longitudinally polarized 
electron beam with degree of polarization $P_L$, for 
$0.1 \leq \Delta \leq 0.5 \  (\Delta = {{E_{\gamma}}\over{E_b}})$ 
and for different beam energies $E_b $, for the electron and positron 
angular range $40^\circ \leq \theta_\pm \leq 140^\circ $. 
No cuts are applied on photon angles. 
The statistical variance (one standard deviation) is given in brackets. 
The parameters are as in Table 3.

\bye